\documentclass[12pt]{article}
\usepackage[utf8]{inputenc}
\usepackage[T1]{fontenc}
\usepackage{lmodern,microtype}
\usepackage[margin=1in]{geometry}
\usepackage{amsmath,amssymb,amsthm,mathtools,bm,comment}
\usepackage{booktabs,array,placeins}
\usepackage{setspace,needspace}
\usepackage[authoryear]{natbib}
\usepackage[colorlinks=true,linkcolor=black,citecolor=black,urlcolor=black,filecolor=black]{hyperref}
\hypersetup{pdftitle={Uniform Inference for Two-Way Clustered Data},
 pdfauthor={Ulrich Hounyo, Jiahao Lin, Xiaojun Song, and Hansheng Wang}}
\allowdisplaybreaks
\newtheorem{theorem}{Theorem}

\newtheorem{remark}{Remark}
\newtheorem{proposition}{Proposition}
\newcommand{\E}{E}
\newcommand{\Var}{\operatorname{Var}}
\newcommand{\op}{O_{\mathrm P}}
\newcommand{\smallop}{o_{\mathrm P}}
\newtheorem{assumption}{Assumption}
\newcommand{\sY}{\sigma_{NT}^{2}}
\newcommand{\sR}{\sigma_{NT,\mathrm R}^{2}}
\newcommand{\sC}{\sigma_{NT,\mathrm C}^{2}}
\newcommand{\sI}{\sigma_{NT,\mathrm I}^{2}}
\newcommand{\hY}{\widehat{\sigma}_{NT}^{2}}
\newcommand{\hR}{\widehat{\sigma}_{NT,\mathrm R}^{2}}
\newcommand{\hC}{\widehat{\sigma}_{NT,\mathrm C}^{2}}
\newcommand{\hI}{\widehat{\sigma}_{NT,\mathrm I}^{2}}
\newcommand{\VF}{V_{NT,\mathrm F}}
\newcommand{\VB}{V_{NT,B}}
\newcommand{\Vone}{V_{NT,1}}
\newcommand{\nuNT}{\nu_{NT}}
\newcommand{\target}{\theta_{NT}}
\newcommand{\arow}[1]{a_{NT,#1}}
\newcommand{\bcol}[1]{b_{NT,#1}}
\newcommand{\resid}[2]{r_{NT,#1#2}}
\newcommand{\btrue}{\bm\beta_{0,NT}}
\newcommand{\dbar}{\bar Y_{NT}}
\newcommand{\calQ}{\mathcal Q}
\newcommand{\calP}{\mathcal P}
\title{Uniform Inference for Two-Way Clustered Data}
\renewcommand{\and}{\unskip\hspace{0.8em}\ignorespaces}
\author{%
{ {Ulrich Hounyo}}\thanks{Department of Economics, University at Albany-State University of New York, Albany, NY 12222, United States. E-mail: \texttt{khounyo@albany.edu}.}
\and
{ {Jiahao Lin}}\thanks{School of Digital Economy and Management, Fuyao University of Science and Technology, Fuzhou, 350109, China. E-mail: \texttt{jhlin@fyust.edu.cn}.} 
\and
{ {Xiaojun Song}}\thanks{Department of Business Statistics and Econometrics, Peking University, Beijing, China. E-mail: \texttt{sxj@gsm.pku.edu.cn}.}
\and 
{ {Hansheng Wang}}\thanks{Guanghua School of Management, Peking University, Beijing, China. E-mail: \texttt{hansheng@gsm.pku.edu.cn}.}}
\date{\today}
\begin{document}
\maketitle

\begin{abstract}
\singlespacing
We establish uniformly asymptotically exact inference for the population mean
of two-way clustered data, allowing the data-generating process to vary with
sample size.
This finding does not contradict the well-known impossibility result of
\citet{menzel2021bootstrap}, which places a fundamental limit on uniform
estimation of the asymptotic distribution of the normalized full-array mean. We estimate the same population mean by averaging the sample means
from random matchings. A closed-form standard error makes the resulting
confidence interval straightforward to implement. The
variance estimator is positive semidefinite 
and the studentized matching estimator is uniformly asymptotically
standard normal. The cost of uniform validity is a slower convergence
rate under degeneracy.  We extend the result to smooth functions of GMM parameters, allowing
the different moment components to exhibit different dependence regimes. In simulations, the proposed confidence intervals achieve
coverage close to the nominal level across the dependence regimes
considered.
\end{abstract}

\newpage

\section{Introduction}\label{sec:intro}

Two-way clustering is widely used in empirical economics to account for
dependence among observations that share a cluster in either of two
dimensions. In a firm panel, for example, outcomes may be correlated
both within a firm over time and across firms in the same period.
The conventional approach estimates a parameter from the full array
and adjusts its standard error for dependence within rows and columns
\citep{cameron2011robust}. Even for the sample mean, however, the
dependence structure can affect the form of the limiting distribution.
When row or column effects dominate, the normalized full-array mean
typically has a Gaussian limit. As these effects weaken, interactions
between rows and columns can produce a non-Gaussian limit.
\citet[Proposition~4.1]{menzel2021bootstrap} shows that the limiting
law of the normalized full-array mean cannot be estimated uniformly
consistently over such sequences. We approach the inference problem
by constructing a different estimator of the same population mean,
for which Gaussian critical values are uniformly valid.

Our estimator averages the means obtained from repeated random
matchings. Taking the dimension with fewer clusters as the rows, each
matching assigns a distinct column to every row, uniformly at random
and independently of the data. No two selected observations share a
row or a column. Under separately exchangeable, dissociated sampling,
they therefore form an i.i.d.\ sample from the marginal distribution.
A single matching permits conventional inference for the population
mean but uses only one observation per row. Additional matchings
recover precision: conditional on the observed array, their means are
independent and centered at the full-array mean. The choice of how
many matchings to use matters because averaging too many can cause
the estimator to inherit the non-Gaussian limit of the full-array mean.

We make four contributions. First, we characterize the limiting distribution of the matching estimator as the number of matchings varies. n exact variance
decomposition separates the sampling variance of the full-array mean from the additional
variance due to matching. Increasing the number of matchings improves precision, but it can also give the full-array error enough weight to produce a non-Gaussian limit. This trade-off motivates letting the number of matchings grow more slowly than the smaller cluster count to recover precision while retaining a Gaussian limit across dependence regimes.

Second, we provide a closed-form variance estimator that makes the
confidence interval straightforward to compute. It combines estimates
of row, column, and intersection variation with nonnegative weights.
The estimate is positive semidefinite in every sample by construction. Inference then
requires only the estimated standard error and a standard normal
critical value, without a bootstrap or a preliminary test of the
dependence regime.

Third, we establish uniformly asymptotically exact inference for the population mean. Under our moment conditions, the studentized matching estimator admits a uniform standard normal approximation whenever the number of matchings is fixed or grows at any rate slower than the smaller cluster count. If the number of matchings grows more slowly than the square root of that count, a uniform bound on the standardized fourth moment suffices for the same conclusion. The approximation holds uniformly over sequences of data-generating processes that may vary with sample size. The strength of clustering in either dimension need not converge, and the two cluster counts may grow at different rates. 

Fourth, we extend the procedure to smooth functions of generalized
method of moments (GMM) parameters, including regression coefficients.
Under two-way clustering, different moments can converge at different
rates. Estimation error in a slowly converging
coefficient can then enter the limiting distribution of a faster one
through a quadratic term \citep[Section~3]{davezies2025analytic}.
We show that the same restriction on the number of matchings makes
this term negligible relative to the standard error of the target.
Applying the variance estimator to the estimated influence function
therefore gives uniformly asymptotically exact inference even when
the moment components follow different dependence regimes.

Our analysis complements the literature on inference based on the
full array. \citet{davezies2021empirical} develop empirical process
theory for exchangeable arrays, and \citet{chiang2023using} give
conditions under which conventional two-way standard errors yield
valid Gaussian inference. \citet{mackinnon2021wild} study wild
bootstrap inference for regression, while \citet*{chiang2025multiway}
construct an empirical likelihood statistic using estimates that leave
out rows or columns. \citet{davezies2025analytic} develop analytic
tests that control size uniformly over their class. Their tests are
asymptotically exact in Gaussian regimes and can be conservative in
non-Gaussian regimes. By changing the estimator, we obtain exact
asymptotic coverage even for laws under which the full-array mean
has a non-Gaussian limit, at a cost in precision that we quantify.

\citet[Supplemental Appendix~C.3]{menzel2021bootstrap} shows that,
when the observation pattern is independent of the data, observing only a
vanishing fraction of intersections can yield a Gaussian limit for
the sample mean. \citet*{chiang2025algorithmic} establish asymptotic normality under
Bernoulli subsampling for each fixed data-generating process.
We select intersections by repeated matching and establish
asymptotically exact coverage uniformly over laws in which row and
column effects may weaken or vanish as the sample size grows.
The role of the number of sampled terms in determining the limiting
distribution also arises in the theory of incomplete $U$-statistics
\citep{janson1984asymptotic,chen2019randomized}.

Section~\ref{sec:setting} introduces the two-way clustering framework
and describes how the dependence structure affects the limiting
distribution of the full-array mean. Section~\ref{sec:multi} presents
the matching estimator, shows how the number of matchings affects its
precision and limiting distribution, and establishes uniformly
asymptotically exact coverage for the resulting interval.
Section~\ref{sec:gmm} extends the results to smooth
functions of GMM parameters. Section~\ref{sec:mc} reports Monte Carlo
evidence on confidence interval coverage and length. Section~6
concludes. The appendix contains the proofs.

\section{Two-way clustering}\label{sec:setting}
\subsection{Sampling framework}

We observe a complete array $Y_{NT}=\{Y_{NT,it}:1\le i\le N,\,
1\le t\le T\}$. Label the smaller dimension as the rows, so that
$2\le N\le T$, and let $N,T\to\infty$ jointly.

\begin{assumption}[Two-way clustered sampling]\label{ass:sampling}
At each $(N,T)$, the observations admit the representation
\begin{equation}
 Y_{NT,it}=f_{NT}(\xi_i,\eta_t,\zeta_{it}).
 \label{eq:array}
\end{equation}
The row shocks $\{\xi_i\}$, column shocks $\{\eta_t\}$, and intersection
shocks $\{\zeta_{it}\}$ form three mutually independent families, with
i.i.d.\ shocks within each family. The marginal mean is
$\target=\E[Y_{NT,it}]$, and the marginal variance satisfies
$\sY=\Var(Y_{NT,it})\in(0,\infty)$.
\end{assumption}

Equation~\eqref{eq:array} is the Aldous-Hoover-Kallenberg representation
for a separately exchangeable, dissociated array
\citep{aldous1981representations,hoover1979relations,kallenberg1989representation}.
Observations may be dependent within a row or column, while collections
of observations with disjoint row and column indices are independent.
Allowing $f_{NT}$ to vary with $(N,T)$ permits both the marginal
distribution and the strength of clustering to change with sample size.
For related work on multiway clustering, see, e.g., 
\citet{davezies2021empirical,davezies2025analytic,menzel2021bootstrap,chiang2023inference,
chiang2024standard,graham2024kernel,chen2024fixed,hounyo2026jackknife,
hounyo2026wild,juodis2026shock,jochmans2026two}.

To describe how each shared shock affects the conditional mean, define
the {row projection} and {column projection} by
\[
 \arow{i}=\E[Y_{NT,it}\mid\xi_i]-\target,\qquad
 \bcol{t}=\E[Y_{NT,it}\mid\eta_t]-\target.
\]
The remaining variation is the {intersection residual}
$\resid{i}{t}$, giving the following decomposition
\begin{equation}
 Y_{NT,it}-\target=\arow{i}+\bcol{t}+\resid{i}{t}.
 \label{eq:projection}
\end{equation}
All three components are centered and mutually uncorrelated. The intersection residual satisfies
$\E[\resid{i}{t}\mid\xi_i]=\E[\resid{i}{t}\mid\eta_t]=0$, although it
can still depend on the interaction between $\xi_i$ and $\eta_t$. Distinct
intersection residuals are uncorrelated.

Averaging \eqref{eq:projection} gives the corresponding decomposition
of the full-array mean $\dbar=(NT)^{-1}\sum_{i,t}Y_{NT,it}$,
\[
 \dbar-\target
 =\frac{1}{N}\sum_{i=1}^N\arow{i}
 +\frac{1}{T}\sum_{t=1}^T\bcol{t}
 +\frac{1}{NT}\sum_{i=1}^N\sum_{t=1}^T\resid{i}{t}.
\]
Let $\sR=\E[\arow{i}^{\,2}]$, $\sC=\E[\bcol{t}^{\,2}]$, and
$\sI=\E[\resid{i}{t}^{\,2}]$, where the subscripts $\mathrm R$,
$\mathrm C$, and $\mathrm I$ denote the row, column, and intersection
components, respectively. By orthogonality,
\begin{equation}
 \sY=\sR+\sC+\sI,\qquad
 \VF=\Var(\dbar)=\frac{\sR}{N}+\frac{\sC}{T}+\frac{\sI}{NT}.
 \label{eq:fullvariance}
\end{equation}

\subsection{Limiting distributions across dependence regimes}

By \eqref{eq:fullvariance}, the row and column projections account for
the following share of the variance of the full-array mean:
\begin{equation}
 \nuNT=\frac{\sR/N+\sC/T}{\VF},\qquad
 1-\nuNT=\frac{\sI/(NT)}{\VF}.
 \label{eq:projection-share}
\end{equation}
We call a sequence \emph{nondegenerate} if $\nuNT\to1$: the projection
averages dominate and have a Gaussian limit after standardization
under the Lindeberg condition for their sum. In the \emph{degenerate}
case, $\nuNT\to0$ and the intersection term dominates.
\emph{Exact degeneracy}, where $\sR=\sC=0$ at every sample size, is
a special case. In the \emph{intermediate} case, $\nuNT\to c\in(0,1)$:
both terms contribute to the variance. These cases describe the
full-array mean; our uniform inference result does not require
$\nuNT$ to converge.

Even under exact degeneracy, the limiting law can be Gaussian or
non-Gaussian. Consider two designs with $\target=0$ and $\sI=\sY=1$,
so that $\VF=1/(NT)$ in both. For $Y_{NT,it}=\zeta_{it}$, with $\zeta_{it}$ standard normal distributed, $\sqrt{NT}\,\dbar$ is standard normal. For
$Y_{NT,it}=\xi_i\eta_t$ with mutually independent standard normal
row and column shocks, $\dbar$ is the product of the two shock averages.
Hence
\[
 \sqrt{NT}\,\dbar
 =(\sqrt N\,\bar\xi_N)(\sqrt T\,\bar\eta_T)
 \overset{d}{=}Z_1Z_2,
\]
where $Z_1$ and $Z_2$ are independent standard normals.

A local shift moves the product design into the intermediate regime. Let
\begin{equation}
 Y_{NT,it}=\xi_i(\eta_t+\delta/\sqrt T).
 \label{eq:product}
\end{equation}
For fixed $\delta\ne0$, $\arow{i}=\delta\xi_i/\sqrt T$, $\bcol{t}=0$, and
$\resid{i}{t}=\xi_i\eta_t$. Although $\sR=\delta^2/T\to0$, the row
projection contributes $\delta^2/(NT)$ to $\VF$, compared with
$1/(NT)$ from the intersection residual. Thus $\nuNT=\delta^2/(1+\delta^2)$, and
\[
 \frac{\dbar-\target}{\sqrt{\VF}}
 \overset{d}{=}\frac{Z_1(Z_2+\delta)}{\sqrt{1+\delta^2}}.
\]
Thus a column shift of order $T^{-1/2}$ changes the law of the
normalized full-array mean. For example, $\delta=0$ and $\delta=1$
yield different limiting distributions, even though the data cannot
consistently distinguish between the two sequences. Consequently, no
estimator of this distribution is uniformly consistent over a class
containing both sequences \citep[Proposition~4.1]{menzel2021bootstrap}.
The next section uses random matching to estimate the same $\target$
with a studentized statistic that admits a uniform Gaussian approximation.

\section{Matching and uniform inference}\label{sec:multi}

\subsection{One matching}\label{sec:one-matching}
A matching selects one observation from each row by assigning a different column to every row. To draw the first matching, start with all $T$ columns available. For row~1, choose uniformly from the $T$ columns. For each subsequent row $i=2,\ldots,N$, choose uniformly from the $T-i+1$ remaining columns. These draws use randomness independent of the array. Write $\pi_1(i)$ for the column assigned to row $i$, leaving the array dimensions $(N,T)$ implicit in the matching notation. By construction,
\begin{equation}
 i\ne j\quad\Longrightarrow\quad\pi_1(i)\ne\pi_1(j).
 \label{eq:distinct}
\end{equation}
Conditional on the draw, the $N$ selected observations come from distinct rows and distinct columns, so they use disjoint triples of shocks in \eqref{eq:array}. They are therefore conditionally i.i.d.\ with the same marginal law as $Y_{NT,11}$.

The matching mean and its sample variance are
\begin{equation}
 M_{NT,1}=\frac{1}{N}\sum_{i=1}^N Y_{NT,i\,\pi_1(i)},\qquad
 S_{NT,1}^2=\frac{1}{N-1}\sum_{i=1}^N(Y_{NT,i\,\pi_1(i)}-M_{NT,1})^2 .
 \label{eq:single}
\end{equation}
For $q\ge2$ and fixed $K<\infty$, let $\calQ_{NT,q}(K)$ denote the
array laws satisfying Assumption~\ref{ass:sampling} and the marginal
moment bound $\E|Y_{NT,11}-\target|^q\le K\sigma_{NT}^q$.
Let $\Phi$ be the standard normal distribution function and $z_p=\Phi^{-1}(p)$. We set a studentized statistic to zero whenever its denominator is zero.

\begin{proposition}[One matching]\label{prop:one}
As $N,T\to\infty$ with $N\le T$,
\[
 \sup_{P\in\calQ_{NT,3}(K)}\sup_{x\in\mathbb R}
 \left|\Pr_P\!\left(\frac{\sqrt N(M_{NT,1}-\target)}{S_{NT,1}}\le x\right)-\Phi(x)\right|\to0 .
\]
\end{proposition}

By Proposition~\ref{prop:one}, the coverage probability of the
one-matching interval $M_{NT,1}\pm z_{1-\alpha/2}S_{NT,1}/\sqrt N$
converges to $1-\alpha$ uniformly over the stated class of laws,
whether or not the projection share $\nuNT$ converges. To assess its precision, write
$\Vone=\Var(M_{NT,1})=\sY/N$. Comparing this variance with
\eqref{eq:fullvariance} gives
\begin{align}
\label{eq:diffV0VF}
 \Vone-\VF
 =\frac{T-N}{NT}\,\sC+\frac{T-1}{NT}\,\sI\ge0.
\end{align}
The row variance cancels because one matching uses every row. On a
square array ($N=T$) it also uses every column, so $\Vone=\VF$ exactly when
$\sI=0$. If $N<T$, equality requires both $\sI=0$ and $\sC=0$.

The precision loss is particularly clear when the intersections are
independent. One matching then averages $N$ observations, while the
full-array mean averages all $NT$, giving $\Vone=T\VF$. A second
matching can use observations omitted by the first. We next examine
how much precision averaging several matching means can recover.

\subsection{Repeated matching and trade-off}
Fix the number of matchings $B=B_{NT}\ge1$ in advance. Draw $B$ independent matchings by the rule in Section~\ref{sec:one-matching}. Write $\pi_b(i)$ for the column assigned to row $i$ in matching $b$. The matching means and their average are
\[
 M_{NT,b}=\frac{1}{N}\sum_{i=1}^N Y_{NT,i\,\pi_b(i)},\qquad
 \widehat\theta_{NT,B}=\frac{1}{B}\sum_{b=1}^B M_{NT,b}.
\]

Conditional on the observed array $Y_{NT}$, the matching means are i.i.d. For each row, every column has probability $1/T$ of being selected, so $\E[M_{NT,b}\mid Y_{NT}]=\dbar$. The error of the matching estimator therefore separates into two terms,
\[
 \widehat\theta_{NT,B}-\target=(\dbar-\target)
    +\frac{1}{B}\sum_{b=1}^B(M_{NT,b}-\dbar).
\]
The first term is the sampling error of the full-array mean. It has
variance $\VF$ from \eqref{eq:fullvariance} and can have a non-Gaussian
limit. The second term is the error from drawing
the matchings. Conditional on $Y_{NT}$, its summands are independent
and centered, so averaging $B$ draws divides the variance of one draw
by $B$. Under the conditions below, this error contributes a Gaussian
term whenever it enters the limit.
\begin{proposition}[Variance of repeated matching]\label{prop:v}
For every $N\le T$ and $B\ge1$, $\E[\widehat\theta_{NT,B}]=\target$ and
\begin{equation}
 \VB:=\Var(\widehat\theta_{NT,B})
 =\underbrace{\VF}_{\text{full-array variance}}
 +\underbrace{\frac{\Vone-\VF}{B}}_{\text{matching variance}}.
 \label{eq:v}
\end{equation}
\end{proposition}

Proposition~\ref{prop:v} holds for any $N\le T$. For the rest of this
subsection, we use square array ($N=T$) to show the trade-off in its simplest form. Every
matching then uses every row and every column exactly once. Only the
average of the selected intersection residuals varies across matchings.
By \eqref{eq:fullvariance},
\eqref{eq:diffV0VF}, and \eqref{eq:v},
\begin{equation}
 \VB=
 \underbrace{\frac{\sR+\sC}{N}+\frac{\sI}{N^2}}_{\text{full-array variance}}
 +\underbrace{\frac{\sI}{NB}\left(1-\frac1N\right)}_{\text{matching variance}}.
 \label{eq:squarevariance}
\end{equation}
The matching variance is proportional to $1/B$, while the full-array
variance is the same for every $B$. By \eqref{eq:squarevariance} and
\eqref{eq:projection-share}, the full-array share of the total variance is
\begin{equation}
 w_{NT,B}=\frac{\VF}{\VB}
 =\frac{B}{B+(N-1)(1-\nuNT)}.
 \label{eq:share}
\end{equation}

Suppose the marginal fourth-moment bound of Proposition~\ref{prop:dense}
holds. In the degenerate and intermediate cases ($\nuNT\to c<1$),
choosing $B=o(N)$ makes the full-array variance share $w_{NT,B}$ vanish,
so the standardized matching estimator has a standard normal limit.
If $B/N\to\lambda\in(0,\infty]$, the full-array error retains a positive
share of the variance and can contribute a non-Gaussian component to
the limit. In the nondegenerate case ($\nuNT\to1$), the standardized
full-array mean is itself asymptotically Gaussian under the projection
Lindeberg condition, so the standardized matching estimator has a
standard normal limit for any deterministic choice of $B$.

The choice of $B$ also determines precision. In the degenerate and
intermediate cases, \eqref{eq:squarevariance} gives
\[
 \VB\asymp \sY\left(\frac{1}{NB}+\frac{1}{N^2}\right).
\]
When $\sigma_{NT}\to\sigma\in(0,\infty)$, the convergence rate is
$\sqrt{NB}$ for $B=o(N)$. Letting $B\to\infty$ within this range
therefore improves precision over one matching while preserving the
Gaussian approximation. At $B=N$, the matching estimator attains the
full-array rate $N$, although $\VB/\VF\to 2-c$. Increasing $B$ further
can still reduce the variance: if $B/N\to\infty$, then $\VB/\VF\to1$,
with no further gain in the convergence rate.

\paragraph{Example 1: Pure product model.}
Consider the pure product model $Y_{NT,it}=\xi_i\eta_t$ on a square array. Here $\nuNT=0$, so
$w_{NT,B}=B/(B+N-1)$, and the standardized full-array mean has the limiting
law of $Z_1Z_2$. Holding the model fixed and varying only $B$ gives
\[
 \frac{\widehat\theta_{NT,B}-\target}{\sqrt{\VB}}
 \ \Rightarrow\
 \begin{cases}
 Z_3,                   & B/N\to0,\\[2pt]
 (Z_1Z_2+Z_3)/\sqrt{2}, & B=N,\\[2pt]
 Z_1Z_2,                & B/N\to\infty,
 \end{cases}
\]
where $Z_1,Z_2,Z_3$ are independent standard normals. The full-array
error  $\dbar-\target$ accounts for none, one half, or all of the limiting variance,
respectively. 

\subsection{The limiting distribution as \texorpdfstring{$B$}{B} varies}\label{sec:Blimits}
In the square-array example, the share $w_{NT,B}$ determines how much
of the non-Gaussian full-array error survives. We now return to the
general setting $N\le T$. When $N<T$, a matching uses every row but
only $N$ of the $T$ columns, so the choice of columns can also
contribute to the matching variance, as \eqref{eq:diffV0VF} shows.
For any array dimensions, \eqref{eq:v} gives the full-array variance share
\begin{equation}
 w_{NT,B}=\frac{\VF}{\VB}
 =\frac{B}{B+\Vone/\VF-1}.
 \label{eq:cost}
\end{equation}
Proposition~\ref{prop:dense} shows how this share weights the two
errors in the limiting distribution.

\begin{proposition}[Limiting law for any number of matchings]\label{prop:dense}
Let $N,T\to\infty$ with $N\le T$. Suppose the array laws belong to
$\calQ_{NT,4}(K)$ for a common $K$, and let $B=B_{NT}\ge1$ be
deterministic. Suppose $w_{NT,B}\to w$.
\begin{enumerate}
\item If $w=0$, then
$(\widehat\theta_{NT,B}-\target)/\sqrt{\VB}\Rightarrow\mathcal N(0,1)$.
\item If $w>0$ and $(\dbar-\target)/\sqrt{\VF}\Rightarrow L_{\mathrm F}$,
then
\[
 \frac{\widehat\theta_{NT,B}-\target}{\sqrt{\VB}}
 \Rightarrow \sqrt w\,L_{\mathrm F}+\sqrt{1-w}\,Z,
\]
where $Z$ is standard normal and independent of $L_{\mathrm F}$.
\end{enumerate}
\end{proposition}

When $w=0$, the matching error alone determines the limit. For
$0<w<1$, an independent Gaussian matching error enters alongside the
full-array error. At $w=1$, the matching estimator inherits the
full-array law. Thus, if the standardized full-array mean is
asymptotically standard normal, so is the standardized matching
estimator for any deterministic sequence $B$, by applying the
proposition along subsequences. In the nondegenerate case
($\nuNT\to1$), a Lindeberg condition on the row and column projection
sums suffices for this conclusion.

\paragraph{The role of the longer dimension.}
In the degenerate and intermediate cases, it is $T$, the number of
columns, that determines the transition. To see why, observe that
\[
 \frac{\Vone}{T\VF}=\frac{\sY}{NT\VF}
 =1-\nuNT+\frac{\sR+\sC}{NT\VF},\qquad
 0\le\frac{\sR+\sC}{NT\VF}\le\frac{\nuNT}{N}.
\]
If $\nuNT\to c<1$, it follows that $(\Vone/\VF-1)/T\to1-c$.
Hence $B=o(T)$ gives $w_{NT,B}\to0$, whereas $B/T\to\infty$ gives
$w_{NT,B}\to1$ and the matching estimator attains the variance of
the full-array mean to first order. When $B/T\to\lambda\in(0,\infty)$,
both errors contribute, with $w_{NT,B}\to\lambda/(\lambda+1-c)$.
Table~\ref{tab:regimes} gives the corresponding rates and limits.
For example, with $B=N$ the pure product model has a non-Gaussian
limit on a square array, but a Gaussian limit if $T/N\to\infty$:
in the latter case $N$ matchings still use only a vanishing fraction
of the $NT$ intersections.

\begin{table}[!htbp]
\centering
\caption{Rates and limiting laws of the matching estimator for $N\le T$}
\label{tab:regimes}
\begin{singlespace}\small
\begin{tabular}{@{}lll@{}}\toprule
Number of matchings & Scale & Limiting distribution \\\midrule
\multicolumn{3}{@{}l}{\emph{Nondegenerate case ($\nuNT\to1$), with a projection Lindeberg condition}}\\[2pt]
Any $B\ge1$ & $1/\sqrt{\VB}$ & $Z$\\
\addlinespace[4pt]
\multicolumn{3}{@{}l}{\emph{Degenerate or intermediate case} ($\nuNT\to c\in[0,1)$)}\\[2pt]
$B/T\to0$ (including fixed $B$) & $\sqrt{NB}/\sigma_{NT}$ & $Z$\\[3pt]
$B/T\to\lambda\in(0,\infty)$ & $\sqrt{NT}/\sigma_{NT}$
  & $L_{\mathrm F}/\sqrt{1-c}+Z/\sqrt\lambda$\\[3pt]
$B/T\to\infty$ & $\sqrt{NT}/\sigma_{NT}$ & $L_{\mathrm F}/\sqrt{1-c}$\\
\bottomrule\end{tabular}
\par\smallskip
\begin{minipage}{.95\linewidth}\footnotesize
\emph{Notes.} The scale multiplies $\widehat\theta_{NT,B}-\target$.
The last two rows assume convergence of the standardized full-array
mean to $L_{\mathrm F}$ as in Proposition~\ref{prop:dense};
$Z$ is standard normal and independent of $L_{\mathrm F}$.
\end{minipage}
\end{singlespace}
\end{table}
Table~\ref{tab:regimes} gives two routes to a Gaussian limit. When $\nuNT\to c<1$, choosing $B=o(T)$ makes the full-array error
negligible. When $\nuNT\to1$, Table~\ref{tab:regimes} gives a Gaussian
limit for any $B$ provided the projection sums satisfy the projection Lindeberg
condition. The $T=N^2$ example in Appendix~\ref{app:boundary} illustrates the
role of the Lindeberg condition in this second case. There, $\nuNT\to1$,
yet the standardized row projection sum converges to a centered Poisson
law. With $B=N$, the share of the matching estimator's variance
attributable to the row projection tends to one half. The standardized
matching estimator therefore has a non-Gaussian limit even though
$B=o(T)$.

\subsection{Variance estimation and uniform Gaussian inference}\label{sec:interval}
We take $B=o(N)$. Under the moment bounds below, this choice yields
asymptotically exact coverage uniformly over the admissible laws,
without requiring $\nuNT$ to converge or imposing a separate
Lindeberg condition on the full-array projection sums.

To accommodate
sequences along which the full-array error remains important, we
estimate the total variance $\VB=(1-1/B)\VF+\sY/(NB)$ from
Proposition~\ref{prop:v}. Let $\bar Y_{NT,i\cdot}$ and
$\bar Y_{NT,\cdot t}$ denote the row and column means. The sample
counterparts of the variance components in \eqref{eq:fullvariance} are
\begin{align}
 \hR&=\frac{1}{N}\sum_i(\bar Y_{NT,i\cdot}-\dbar)^2,&
 \hC&=\frac{1}{T}\sum_t(\bar Y_{NT,\cdot t}-\dbar)^2,\notag\\
 \widehat{r}_{NT,it}&=Y_{NT,it}-\bar Y_{NT,i\cdot}-\bar Y_{NT,\cdot t}+\dbar,&
 \hI&=\frac{1}{NT}\sum_{i,t} \widehat{r}_{NT,it}^2 .
 \label{eq:components}
\end{align}
The marginal variance estimate $\hY=(NT)^{-1}\sum_{i,t}(Y_{NT,it}-\dbar)^2$ equals $\hR+\hC+\hI$. Substitution in \eqref{eq:v} gives variance estimator
\begin{equation}
 \widehat V_{NT,B}=
 \left(1-\frac{1}{B}\right)
 \left(\frac{\hR}{N}+\frac{\hC}{T}+\frac{\hI}{NT}\right)
 +\frac{\hY}{NB}.
 \label{eq:se}
\end{equation}
The estimator $\widehat V_{NT,B}$ is positive semidefinite by construction in every sample.
The standard two-way cluster-robust variance estimator for full-array mean subtracts the
intersection component and can therefore yield a negative variance
estimate \citep{cameron2011robust,chiang2024standard}.

The matching estimator uses the selected observations, while its standard error $\sqrt{\widehat V_{NT,B}}$ uses the full array. The resulting confidence interval is
\begin{equation}
 I_{NT,B}(1-\alpha)=
 \left[\widehat\theta_{NT,B}-z_{1-\alpha/2}\sqrt{\widehat V_{NT,B}},\;
       \widehat\theta_{NT,B}+z_{1-\alpha/2}\sqrt{\widehat V_{NT,B}}\right].
 \label{eq:interval}
\end{equation}
For any deterministic $B$ such that $w_{NT,B}\to0$, the proof of
Proposition~\ref{prop:dense} also gives $\widehat V_{NT,B}/\VB\to1$
in probability under the marginal fourth-moment bound alone.
Studentization therefore preserves the Gaussian limit along those sequences.
For uniform coverage we must
also allow the full-array error to retain a positive variance share.
Every matching includes every row, so an unusually large row projection
can affect every matching mean. We control such projections through
the following moment bounds.

\begin{assumption}[Moment bounds]\label{ass:moments}
For a constant $K$ common to all laws and array sizes under consideration,
 $\E[\arow{1}^{\,4}]\le K\sY\sR,$ and 
$\E[\bcol{1}^{\,4}]\le K\sY\sC$.
\end{assumption}

Let $\calP_{NT}(K)$ denote the subclass of $\calQ_{NT,4}(K)$ that
also satisfies the row and column projection bounds in Assumption \ref{ass:moments}.
The moment bounds allow either projection variance to approach zero.

\begin{theorem}[Uniform Gaussian inference]\label{th:main}
As $N,T\to\infty$ with $N\le T$, let $B=B_{NT}\ge1$ be deterministic with $B=o(N)$. For every $\epsilon>0$,
\begin{align}
 &\sup_{P\in\calP_{NT}(K)}\Pr_P\!\left(\left|\frac{\widehat V_{NT,B}}{\VB}-1\right|>\epsilon\right)\to0,\label{eq:cons}\\
 &\sup_{P\in\calP_{NT}(K)}\sup_{x\in\mathbb R}
 \left|\Pr_P\!\left(\frac{\widehat\theta_{NT,B}-\target}{\sqrt{\widehat V_{NT,B}}}\le x\right)-\Phi(x)\right|\to0 .\label{eq:clt}
\end{align}
If $B=o(\sqrt N)$, \eqref{eq:cons} and \eqref{eq:clt} also hold uniformly over
$\calQ_{NT,4}(K)$ without the projection bounds in Assumption \ref{ass:moments}.
\end{theorem}

For $B=o(N)$, the interval in \eqref{eq:interval} has uniformly
asymptotically exact coverage over $\calP_{NT}(K)$. The projection
share $\nuNT$ may vary without converging. The theorem also requires
no separate Lindeberg condition of the kind used in
Table~\ref{tab:regimes} to obtain a Gaussian limit for unrestricted
$B$ when $\nuNT\to1$.\footnote{When $\nuNT\to1$, the moment bounds
defining $\calP_{NT}(K)$ imply the projection Lindeberg condition if
$T/N$ is bounded. It can fail when $T/N\to\infty$ even under those
bounds. See Appendix~\ref{app:boundary}.}

\begin{remark}Theorem~\ref{th:main} permits any deterministic sequence $B=o(N)$.
On a square array, one can therefore choose $B$ so that the matching
estimator attains the convergence rate of the full-array mean up to
an arbitrarily slowly diverging factor. For example, under
exact degeneracy,
$B=\lfloor N/\log N\rfloor$ gives a standard error asymptotic to
$\sigma_{NT}\sqrt{\log N}/N$, compared with $\sigma_{NT}/N$ for the
full-array mean. The main cost of uniform inference in this case is therefore
an additional factor of $(\log N)^{-1/2}$ in the convergence rate.
\end{remark}

The weaker moment condition in Theorem~\ref{th:main} places a separate restriction on $B$. With only a bounded standardized marginal fourth moment, an influential row can be rare enough that its sample count has a Poisson limit. Appendix~\ref{app:boundary} constructs a counterexample for every sequence with $B=o(N)$ for which $B/\sqrt N$ fails to converge to zero. Thus $B=o(\sqrt N)$ is sharp for the uniform Gaussian approximation under a marginal fourth-moment bound alone.

\section{Extension to generalized method of moments}\label{sec:gmm}
The population mean solves the moment condition $\E(Y_{NT,it}-\target)=0$. GMM replaces this one equation by a vector of equations. At the true parameter, each row-column intersection contributes a vector of moments. The components of this vector may exhibit different dependence regimes. We show that the Gaussian inference
developed in Section~\ref{sec:multi} extends to smooth functions of
the GMM parameter even when the sample moments converge at different
rates.

Let $\bm X_{NT,it}=f_{NT}(\xi_i,\eta_t,\zeta_{it})$ follow the sampling model in \eqref{eq:array}. Let $\bm m_{NT}(\bm X_{NT,it},\bm\beta)$ be a vector of $r$ moment functions. Suppose a $d$-dimensional parameter $\btrue$ satisfies $\E[\bm m_{NT}(\bm X_{NT,it},\btrue)]=\bm 0$, where $r\ge d$ and both dimensions are fixed. We consider inference on a scalar quantity $\ell(\btrue)$. In the GMM framework of \citet{hansen1982large}, the population moments identify $\btrue$. We replace their usual sample average by the matching average. Using the same $B$ matchings at every candidate value $\bm\beta$, compute
\begin{align}
 \widehat{\bm m}_{NT,B}(\bm\beta)
  &=\frac{1}{NB}\sum_{b=1}^{B}\sum_{i=1}^{N}
      \bm m_{NT}(\bm X_{NT,i\,\pi_b(i)},\bm\beta),\notag\\
 \widehat{\bm\beta}_{NT,B}
  &\in\arg\min_{\bm\beta\in\Theta}
      \widehat{\bm m}_{NT,B}(\bm\beta)^\top\widehat{\bm W}_{NT}\,
      \widehat{\bm m}_{NT,B}(\bm\beta).
 \label{eq:gmm-estimator}
\end{align}
The weight $\widehat{\bm W}_{NT}$ is held fixed in this minimization. Exactly identified and overidentified models use the same matching average.

Let $\dot{\bm m}_{NT}$ be the $r\times d$ derivative of the moment vector and define
\[
 \bm D_{NT}=\E\dot{\bm m}_{NT}(\bm X_{NT,11},\btrue),\qquad
 \bm A_{NT}=(\bm D_{NT}^\top\bm W_{NT}\bm D_{NT})^{-1}\bm D_{NT}^\top\bm W_{NT},
\]
where $\bm W_{NT}$ is the population weight. Linearizing the moment conditions gives the contribution of intersection $(i,t)$ to the estimation error in $\ell(\widehat{\bm\beta}_{NT,B})$,
\begin{equation}
 u_{NT,it}=-\nabla\ell(\btrue)^\top\bm A_{NT}\,
               \bm m_{NT}(\bm X_{NT,it},\btrue).
 \label{eq:gmm-score}
\end{equation}
For the population mean, this contribution equals $Y_{NT,it}-\target$. For a general GMM parameter, we apply the variance estimator in \eqref{eq:se} to the scalar contributions in \eqref{eq:gmm-score}. To obtain their fitted values $\widehat u_{NT,it}$, use $\widehat{\bm\beta}_{NT,B}$, the matching average of $\dot{\bm m}_{NT}$ evaluated at $\widehat{\bm\beta}_{NT,B}$, and $\widehat{\bm W}_{NT}$ in place of $\btrue$, $\bm D_{NT}$, and $\bm W_{NT}$, respectively. Denote the fitted derivative by $\widehat{\bm D}_{NT,B}$. Writing $\widehat V_{NT,B}(x)$ for \eqref{eq:se} applied to a scalar array $x$, the resulting standard error is $\sqrt{\widehat V_{NT,B}(\widehat u_{NT})}$.

\begin{assumption}[GMM regularity]\label{ass:gmm}
The following conditions hold uniformly over array sizes and laws.
\begin{enumerate}
\item The parameter space $\Theta$ is fixed, compact, and convex.
The distance from $\btrue$ to the boundary of $\Theta$ and the smallest
singular value of $\bm D_{NT}$ are both bounded away from zero.
For every $\epsilon>0$, there is a $c_\epsilon>0$ such that
$\|\E\bm m_{NT}(\bm X_{NT,11},\bm\beta)\|\ge c_\epsilon$ for all
$\bm\beta\in\Theta$ with $\|\bm\beta-\btrue\|\ge\epsilon$.

\item For every $\bm x$, each $m_{NT,j}(\bm x,\cdot)$,
$j=1,\ldots,r$, is twice continuously differentiable on a neighborhood
of $\Theta$. On $\Theta$, the moment functions and their first two
derivatives are bounded in norm by $F_{NT}(\bm x)$, with
$\E[F_{NT}(\bm X_{NT,11})^2]\le K$. The gradient and Hessian of
$\ell$ are continuous and bounded on $\Theta$.

\item The weight $\widehat{\bm W}_{NT}$ is symmetric positive definite
and satisfies $\|\widehat{\bm W}_{NT}-\bm W_{NT}\|
=O_{\mathrm P}(N^{-1/2})$, where $\bm W_{NT}$ is deterministic and
symmetric with eigenvalues bounded above and away from zero.
\end{enumerate}
\end{assumption}

Assumption~\ref{ass:gmm} ensures that the population moments uniformly
identify $\btrue$ and that their Jacobian has full column rank at
$\btrue$. The smoothness and moment bounds justify linearizing the GMM
estimator and its scalar target, while the weighting condition permits
either a fixed weight or a weight estimated at the $\sqrt N$ rate.

For $q\ge2$ and fixed constants $K$ and $\underline v>0$, let
$\calQ_{NT,\mathrm{GMM},q}(K)$ denote the class of GMM laws $P$
satisfying Assumption~\ref{ass:gmm} such that
$\Var_P(u_{NT,11})\ge\underline v$ and the law of the influence array
$u_{NT}$ belongs to $\calQ_{NT,q}(K)$. Let
$\calP_{NT,\mathrm{GMM}}(K)$ be the subclass of
$\calQ_{NT,\mathrm{GMM},4}(K)$ for which the row and column projections
of $u_{NT}$ also satisfy Assumption \ref{ass:moments}.

\begin{theorem}[Matching GMM]\label{th:gmm}
Let $N,T\to\infty$ with $N\le T$, and let $B=o(N)$ be deterministic.
Then
\begin{equation}
 \sup_{P\in\calP_{NT,\mathrm{GMM}}(K)}\sup_{x\in\mathbb R}
 \left|
 \Pr_P\!\left(
   \frac{\ell(\widehat{\bm\beta}_{NT,B})-\ell(\btrue)}
        {\sqrt{\widehat V_{NT,B}(\widehat u_{NT})}}
   \le x
 \right)-\Phi(x)
 \right|\to0.
 \label{eq:gmm-clt}
\end{equation}
If $B=o(\sqrt N)$, the convergence in \eqref{eq:gmm-clt} holds
uniformly over the larger class $\calQ_{NT,\mathrm{GMM},4}(K)$.
\end{theorem}

The lower variance bound $\underline{v}$ concerns variation in the contribution from a single intersection $u_{NT,11}$. Both its projection variances may equal zero.  Appendix~\ref{app:gmm} also shows that, when the stated bounds hold for all unit linear contrasts, the joint Wald statistic for $\btrue$ has a uniform $\chi^2_d$ limit.

\begin{remark}
For nonlinear GMM, a quadratic term in the estimation error can enter the limiting distribution under two-way clustering \citep{davezies2025analytic}. In the present construction the estimation remainder is $O_{\mathrm P}(N^{-1})$, whereas the standard deviation of the matching average of $u_{NT}$ is at least $\sqrt{\Var(u_{NT,11})/(NB)}$. Thus $B=o(N)$ makes the remainder negligible even when the quantity of interest converges faster than other components of $\widehat{\bm\beta}_{NT,B}$. 
\end{remark}

\paragraph{Example 2: GMM moments with different dependence regimes.}
On a square array, observe $\bm X_{NT,it}=(x_{1,i},x_{2,it})^\top$,
where $\{x_{1,i}\}_{i=1}^N$ and $\{x_{2,it}\}_{i,t=1}^N$ are mutually
independent standard normal draws.  Define the moment function by
\(
 \bm m_{NT}(\bm X_{NT,it},\bm\beta)
 =\begin{pmatrix}x_{1,i}-\beta_1\\ x_{2,it}+\beta_1^2-\beta_2\end{pmatrix}.
\)
The first population moment gives $\beta_1=\E[x_{1,i}]=0$.
Substituting into the second gives
$\beta_2=\E[x_{2,it}]+\beta_1^2=0$. Thus $\btrue=\bm 0$, and we take
$\ell(\bm\beta)=\beta_2$.

Let $\bar x_1=N^{-1}\sum_i x_{1,i}$ and
$\bar x_2=N^{-2}\sum_{i,t}x_{2,it}$. The full-array moment equations
give $\bar x_1$ for the first coefficient and $\bar x_2+\bar x_1^2$
for the second. Since $N\bar x_2$ and $\sqrt N\,\bar x_1$ are
independent standard normals,
\[
 N\widehat\beta_{NT,\mathrm F,2}
 = N\bar x_2+(\sqrt N\,\bar x_1)^2+o_{\mathrm P}(1)
 \Rightarrow Z_0+Z_1^2,
\]
where $Z_0$ and $Z_1$ are independent standard normals. This illustrates a key
difficulty of the full-array estimator identified by \citet[Section~3]{davezies2025analytic}:
 the estimation error in the slower first coefficient enters the limiting law of the faster second coefficient.

Every matching uses all $N$ rows, so the first moment equation still
gives $\bar x_1$. In the second equation, $\bar x_2$ is replaced by
\(
 \bar x_{2,B}=\frac{1}{NB}\sum_{b=1}^B\sum_{i=1}^N x_{2,i\,\pi_b(i)}.
\)
For $B=o(N)$, Table~\ref{tab:regimes} gives
$\sqrt{NB}\,\bar x_{2,B}\Rightarrow\mathcal N(0,1)$. The quadratic
term is now negligible because $\sqrt{B/N}\to0$. Hence
\[
 \sqrt{NB}\,\widehat\beta_{NT,B,2}
 = \sqrt{NB}\,\bar x_{2,B}
   +\sqrt{\frac{B}{N}}\,(\sqrt N\,\bar x_1)^2+o_{\mathrm P}(1)
 \Rightarrow \mathcal N(0,1).
\]
The Gaussian approximation remains valid even when different components
of the GMM moment vector exhibit different dependence regimes.

\section{Monte Carlo evidence for the mean}\label{sec:mc}
We return to the sample mean to measure how much repeated matching shortens an interval and how closely each method attains nominal coverage. Each design has $N=T$ and $\target=0$. Let $\xi_i$, $\eta_t$, and $\zeta_{it}$ be mutually independent standard normal shocks, independent across indices within each family. The four designs specify the outcome at each intersection as follows.
\[
\begin{array}{lll}
\text{Independent intersections} &Y_{NT,it}=\zeta_{it},&
\text{Additive effects} \quad Y_{NT,it}=\xi_i+\eta_t,\\[3pt]
\text{Pure product} &Y_{NT,it}=\xi_i\eta_t,&
\text{Local product} \quad Y_{NT,it}=\xi_i(\eta_t+N^{-1/2}).
\end{array}
\]
Independent intersections and additive effects give two Gaussian benchmarks with quite different costs of matching.  The two product designs have the same intersection variance, $\sI=1$, and zero column projection variance. Their row projection variances are $\sR=0$ for the pure product and $\sR=1/N$ for the local product. The pure product is therefore exactly degenerate, whereas the local product is intermediate, with $\nuNT=1/2$. They are the $\delta=0$ and $\delta=1$ cases of \eqref{eq:product}, the local comparison in \citet[Proposition~4.1]{menzel2021bootstrap}.

For each array we construct five nominal 95\% confidence intervals. For one matching we use the ordinary sample standard error of the $N$ selected observations. For repeated matching we use \eqref{eq:interval} with $B=\lfloor N^{1/3}\rfloor$ and $B=\lfloor N/\log N\rfloor$. Both choices satisfy $B=o(N)$, and the cube-root choice also satisfies $B=o(\sqrt N)$. All three matching intervals use the standard normal critical value and the same sequence of matching draws, taking the first $B$ draws for each choice.

The other two intervals use the full-array mean. For the two-way
cluster-robust interval of \citet{cameron2011robust} (CGM), the
degrees-of-freedom-adjusted variance estimate on the square array is
\[
 \widehat V_{\mathrm{CGM}}
 =\frac{\hR+\hC}{N-1}-\frac{\hY}{N^2-1}.
\]
The interval is $\dbar\pm z_{0.975}\sqrt{\max\{0,\widehat V_{\mathrm{CGM}}\}}$.
A nonpositive variance estimate yields a zero-length interval centered
at $\dbar$. We include these replications in both coverage and mean length.

The bootstrap interval uses the model-selection procedure (BS-S) of
\citet[Definition~3.1]{menzel2021bootstrap} and its equal-tailed,
nonstudentized interval (REG). Table~\ref{tab:mc} uses the selection
threshold $\log(N)/N$ for both corrected projection variances in
every design.\footnote{We resample rows and columns independently with
replacement and apply crossed Mammen multipliers to the resampled
double-demeaned residuals. If $q_j^*$ denotes the $j$th ordered centered
bootstrap mean among the 999 draws, the interval is
$[\dbar-q_{975}^*,\,\dbar-q_{25}^*]$. Menzel also proposes a
conservative procedure for uniform size control.}

\begin{table}[!htbp]
\centering
\caption{Coverage and length of nominal 95\% confidence intervals}
\label{tab:mc}
\begin{singlespace}
\footnotesize
\setlength{\tabcolsep}{2.5pt}
\renewcommand{\arraystretch}{1.16}
\begin{tabular*}{\linewidth}{@{}l@{\extracolsep{\fill}}r*{5}{r}*{5}{r}@{}}
\toprule
 & & \multicolumn{5}{c}{Coverage (\%)} & \multicolumn{5}{c}{Mean length}\\
\cmidrule(lr){3-7}\cmidrule(l){8-12}
Design & $N$ & One & $N^{1/3}$ & $\tfrac{N}{\log N}$ & CGM & BS-S
             & One & $N^{1/3}$ & $\tfrac{N}{\log N}$ & CGM & BS-S\\
\midrule
Independent & 25 & 93.73 & 95.60 & 97.44 & 92.29 & 94.11 & 0.777 & 0.584 & 0.384 & 0.153 & 0.154 \\
 & 50 & 94.50 & 95.91 & 97.68 & 93.69 & 94.80 & 0.552 & 0.338 & 0.205 & 0.078 & 0.078 \\
 & 100 & 94.98 & 95.71 & 97.79 & 94.23 & 94.64 & 0.390 & 0.205 & 0.108 & 0.039 & 0.039 \\
\addlinespace[3pt]
Additive & 25 & 94.03 & 94.08 & 94.08 & 94.11 & 94.16 & 1.095 & 1.080 & 1.080 & 1.081 & 1.084 \\
 & 50 & 94.70 & 94.49 & 94.49 & 94.50 & 94.40 & 0.780 & 0.773 & 0.773 & 0.773 & 0.776 \\
 & 100 & 95.17 & 94.95 & 94.95 & 94.96 & 94.97 & 0.553 & 0.551 & 0.551 & 0.551 & 0.553 \\
\addlinespace[3pt]
Pure product & 25 & 94.42 & 95.95 & 97.85 & 61.86 & 95.98 & 0.759 & 0.572 & 0.374 & 0.119 & 0.181 \\
 & 50 & 94.58 & 95.97 & 97.73 & 61.21 & 95.75 & 0.542 & 0.334 & 0.202 & 0.060 & 0.091 \\
 & 100 & 95.18 & 95.86 & 97.86 & 59.67 & 95.92 & 0.388 & 0.203 & 0.107 & 0.030 & 0.045 \\
\addlinespace[3pt]
Local product & 25 & 94.47 & 95.83 & 98.09 & 74.12 & 93.27 & 0.774 & 0.591 & 0.401 & 0.174 & 0.206 \\
 & 50 & 94.54 & 95.77 & 97.98 & 73.08 & 92.84 & 0.548 & 0.343 & 0.215 & 0.088 & 0.103 \\
 & 100 & 95.08 & 95.73 & 97.91 & 72.78 & 92.21 & 0.390 & 0.207 & 0.113 & 0.044 & 0.050 \\
\bottomrule
\end{tabular*}
\par\smallskip
\begin{minipage}{\linewidth}\footnotesize
\emph{Notes.} One uses a single matching. The repeated matching uses the $N^{1/3}$ rule and the $N/\log N$ rule, respectively. CGM and BS-S use the full-array mean. Each design and size uses 10,000 Monte Carlo replications, with 999 bootstrap draws per replication for BS-S. 
\end{minipage}
\end{singlespace}
\end{table}

Coverage with one matching approaches the nominal 95\% level across
the sample sizes in Table~\ref{tab:mc}, in line with
Proposition~\ref{prop:one}. Repeated matching recovers much of the
precision lost by using only one matching. Under independence and in
both product designs, the cube-root rule gives intervals about half
as long as those from one matching at $N=100$, with coverage close to
nominal at all three sample sizes. Increasing $B$ to
$\lfloor N/\log N\rfloor$ shortens the intervals further but leads to
greater overcoverage. Under additive effects, the two rules give
identical intervals because every matching averages the same row and
column effects.

Adding the local row effect changes coverage little for either
repeated-matching rule, as the full-array error is asymptotically
negligible in both product designs. BS-S gives shorter intervals in
both designs and achieves close to nominal coverage under the pure
product. Under the local product, its coverage still appears reasonably
close to nominal in these samples, although it falls to about 92\% at
$N=100$. The row projection accounts for half the variance of the
full-array mean, yet its variance $1/N$ lies below the BS-S selection
threshold $\log(N)/N$. Consequently, BS-S omits the row effect with
probability tending to one and continues to approximate the pure-product
law, so its limiting coverage under the local product is below 95\%.
CGM undercovers more severely in both designs and frequently produces
zero-length intervals.

Based on these results, we recommend $B=\lfloor N^{1/3}\rfloor$ as
a default. It delivers a substantial gain in precision over one
matching while keeping finite-sample coverage close to nominal.
When shorter intervals are the priority, $B=\lfloor N/\log N\rfloor$
is a reasonable choice, with the trade-off of more conservative
coverage in the independent and product designs.

\section{Conclusion}
A single matching produces an i.i.d. sample because every selected observation has a different row index and a different column index.  We average matching means and use a
closed-form standard error with the usual Gaussian critical value.
Coverage is uniformly asymptotically exact even when the limiting law
of the normalized full-array mean cannot be estimated uniformly. Repeated matching recovers much of the
precision of the full array. The same variance formula yields inference for smooth functions of GMM
parameters. With $B=o(N)$, a nonlinear remainder is negligible on the matching
scale even when the moment components exhibit different dependence
regimes. In such settings, the remainder can contribute to the limiting
distribution of the full-array GMM estimator \citep{davezies2025analytic}.

\appendix

\section{One matching and the exact variance}\label{app:basic}
\begin{proof}[Proof of Proposition~\ref{prop:one}]
Condition on the random matching $\pi_1$. By \eqref{eq:distinct}, the selected observations use disjoint triples of shocks. Since the draw is independent of the array, for any $z_1,\ldots,z_N$,
\[
 \Pr\!\left(\bigcap_{i=1}^{N}\{Y_{NT,i\,\pi_1(i)}\le z_i\}
             \,\middle|\,\pi_1\right)
 =\prod_{i=1}^{N}\Pr(Y_{NT,11}\le z_i).
\]
The right-hand side is the same for every draw, so the product law also holds unconditionally.
Write $X_{NT,i}=(Y_{NT,i\,\pi_1(i)}-\target)/\sigma_{NT}$. Along any sequence of laws in $\calQ_{NT,3}(K)$, the $X_{NT,i}$ within each sample are i.i.d. with mean zero, variance one, and $\E|X_{NT,i}|^3\le K$. The Berry-Esseen inequality gives a standard normal approximation for $N^{-1/2}\sum_i X_{NT,i}$ with error at most $CK/\sqrt N$. For sample variance consistency, truncate $X_{NT,i}^2$ at a fixed $L^2$. The truncated sample average concentrates around its expectation by Chebyshev's inequality, and $\E[X_{NT,i}^2\mathbf 1\{|X_{NT,i}|>L\}]\le K/L$. Letting $L\to\infty$ gives $N^{-1}\sum_i X_{NT,i}^2\to 1$ in probability. Since $\E[\bar X_{NT}^2]=1/N$,
\[
 \frac{S_{NT,1}^2}{\sY}
 =\frac{N}{N-1}\left(\frac1N\sum_i X_{NT,i}^2-\bar X_{NT}^2\right)
 \xrightarrow{\mathrm P}1 .
\] In particular, the probability of a zero denominator vanishes. Slutsky's theorem and continuity of $\Phi$ give convergence in Kolmogorov distance along every admissible sequence, proving uniformity. For a direct bound on the studentized statistic, see \citet{bentkus1996berry}.
\end{proof}

\begin{proof}[Proof of Proposition~\ref{prop:v}]
Each column is equally likely to be assigned to row $i$, so
$\E[M_{NT,b}\mid Y_{NT}]=N^{-1}\sum_i T^{-1}\sum_tY_{NT,it}=\dbar$.
Thus $\E\widehat\theta_{NT,B}=\E\dbar=\target$. Conditional independence of the $B$ draws and the law of total variance give
\begin{align*}
 \Var(\widehat\theta_{NT,B})
  &=\Var(\dbar)+\frac{1}{B}\E\Var(M_{NT,1}\mid Y_{NT}),\\
 \E\Var(M_{NT,1}\mid Y_{NT})
  &=\Var(M_{NT,1})-\Var(\dbar).
\end{align*}
Substitution gives \eqref{eq:v}.
\end{proof}

\section{Proof of Theorem~\ref{th:main}}\label{app:proof}
Fix any sequence of laws in $\calP_{NT}(K)$ as $N,T\to\infty$ with $N\le T$. Both conclusions are invariant to centering each array at $\target$ and dividing its outcomes by $\sigma_{NT}$. We work with the standardized array throughout this proof, so $\target=0$ and $\sY=1$ at each $(N,T)$. Constants in the bounds below depend only on $K$. From \eqref{eq:v},
\begin{equation}
 \VB\ge\VF,\qquad \VB\ge\frac{1}{NB},\qquad
 \VF=\frac{\sR}{N}+\frac{\sC}{T}+\frac{\sI}{NT}.
 \label{eq:floor}
\end{equation}
Write $\bar a_{NT}$, $\bar b_{NT}$ and $\bar r_{NT}$ for the sample averages of the row projections, column projections and residuals, respectively.

\paragraph{Variance estimation.}
Distinct residuals are uncorrelated, so their row, column and full-array averages satisfy
\[
 \E\bar r_{NT,i\cdot}^2=\frac{\sI}{T},\qquad
 \E\bar r_{NT,\cdot t}^2=\frac{\sI}{N},\qquad
 \E\bar r_{NT}^2=\frac{\sI}{NT}.
\]
To bound the row variance estimate, write its centered row means in terms of the true projections and residuals:
\[
 \hR=\frac1N\sum_i
 \bigl[(\arow{i}-\bar a_{NT})+(\bar r_{NT,i\cdot}-\bar r_{NT})\bigr]^2 .
\]
The mean square of the residual contrasts is bounded by
$N^{-1}\sum_i\bar r_{NT,i\cdot}^2=\op(\sI/T)$. The mean square of the projection contrasts is $\op(\sR)$, so Cauchy-Schwarz bounds the cross term by $\op(\sqrt{\sR\,\sI/T})$. For the projection term, independence and Assumption~\ref{ass:moments} give
\[
 \Var\!\left(N^{-1}\sum_i \arow{i}^2\right)\le \frac{K\sR}{N},
 \qquad \E\bar a_{NT}^2=\frac{\sR}{N}.
\]
Applying the same argument to the columns gives the bounds
\begin{align}
 |\hR-\sR|&=\op\!\left(\sqrt{\sR/N}+\sqrt{\sR\,\sI/T}+\sI/T\right),\notag\\
 |\hC-\sC|&=\op\!\left(\sqrt{\sC/T}+\sqrt{\sC\,\sI/N}+\sI/N\right).
 \label{eq:rowrate}
\end{align}
By Jensen's inequality and the marginal fourth-moment bound, $\E \resid{1}{1}^4$ is bounded uniformly. The squares $\resid{i}{t}^2$ and $\resid{i'}{t'}^2$ are independent when $i\ne i'$ and $t\ne t'$. Counting pairs with a common index gives $\Var((NT)^{-1}\sum_{i,t}\resid{i}{t}^2)=O(1/N+1/T)$. Double demeaning the outcome has the same effect as double demeaning the residual. Thus
\[
 0\le \frac1{NT}\sum_{i,t}\resid{i}{t}^2-\hI
 =\frac1N\sum_i\bar r_{NT,i\cdot}^2+\frac1T\sum_t\bar r_{NT,\cdot t}^2-\bar r_{NT}^2
 =\op(1/N).
\]
Consequently, $\hI-\sI=\op(N^{-1/2})$. The same pair count applies to $Y_{NT,it}^2$, and $\E[\dbar^2]=\VF\le1/N$. Thus $\hY-1=\op(N^{-1/2})$ as well.

By \eqref{eq:floor}, $N\VB\ge\max\{\sR,1/B\}$ and $T\VB\ge\max\{\sC,1/B\}$. In particular $\sqrt{x}/\max\{x,1/B\}\le\sqrt B$ for all $x\ge0$. Dividing the first and second bounds in \eqref{eq:rowrate} by $N\VB$ and $T\VB$, respectively, therefore gives
\[
 \frac{|\hR-\sR|}{N\VB}+\frac{|\hC-\sC|}{T\VB}
 =\op\!\left(\sqrt{B/N}+B/N\right).
\]
The estimation errors for $\sI$ and $\sY$ in \eqref{eq:se}, divided by $\VB$, are $o_{\mathrm P}(1)$ since $NT\VB\ge T/B$ and $NB\VB\ge1$. This proves \eqref{eq:cons}.

\paragraph{Variation in the full-array mean.}
The identity $\dbar=\bar a_{NT}+\bar b_{NT}+\bar r_{NT}$ separates its independent projection sums from the residual average. By \eqref{eq:floor}, $\E\bar r_{NT}^2/\VB\le B/T\to0$. Write $V_{NT,\mathrm P}=\sR/N+\sC/T=\nuNT\VF$. The sum of fourth moments of the summands of $(\bar a_{NT}+\bar b_{NT})/\sqrt{\VB}$ satisfies
\[
 \frac{1}{\VB^2}\left(\frac{\E \arow{1}^4}{N^3}+\frac{\E \bcol{1}^4}{T^3}\right)
 \le\frac{K V_{NT,\mathrm P}}{N^2\VB^2}\le\frac{KB}{N}\to0 .
\]
From any subsequence take a further subsequence with $V_{NT,\mathrm P}/\VB\to\ell\in[0,1]$. If $\ell>0$, Lyapunov's theorem gives $\dbar/\sqrt{\VB}\Rightarrow \mathcal N(0,\ell)$. If $\ell=0$, the normalized full-array mean converges to zero in mean square.

\paragraph{Variation from drawing the matchings.}
Put $D_{NT,b}=M_{NT,b}-\dbar$. Given the full array $Y_{NT}$, the $D_{NT,b}$ are i.i.d. with conditional mean zero. The conditional variance has the observable form
\begin{equation}
 \omega_{NT}^2(Y_{NT}):=N\Var(D_{NT,1}\mid Y_{NT})
   =\frac{T}{T-1}\left(\hI+(1-N/T)\hC\right).
 \label{eq:H}
\end{equation}
To verify \eqref{eq:H}, pad the $N$ row-centered observations in each column with $T-N$ zeros to obtain a $T\times T$ matrix $A$. Write $g_t=\bar Y_{NT,\cdot t}-\dbar$. The mean of column $t$ of $A$ is $(N/T)g_t$. Let $C$ be $A$ after subtracting these column means, so all its row and column sums are zero. Since $(NT)^{-1}\sum_{i,t\le T}A_{it}^2=\hI+\hC$ and $T^{-1}\sum_t g_t^2=\hC$,
\[
 \frac{1}{NT}\sum_{i,t\le T}C_{it}^2
 =\hI+\hC-\frac{1}{N}\sum_{t=1}^{T}\left(\frac{N}{T}g_t\right)^2
 =\hI+(1-N/T)\hC .
\]
Complete the matching to a uniform permutation of the $T$ columns. Its sum over $C$ equals $ND_{NT,1}$, since the subtracted column means sum to zero. Put $S_C=\sum_{i,t\le T}C_{it}^2$. The zero row and column sums give $\sum_{j\ne i,\,s\ne t}C_{js}=C_{it}$. A single permutation assignment has probability $1/T$, and a specified pair in distinct rows and columns has probability $1/[T(T-1)]$. Hence
\[
 \Var(ND_{NT,1}\mid Y_{NT})
 =\frac{S_C}{T}
  +\frac{\sum_{i,t\le T} C_{it}\sum_{j\ne i,\,s\ne t}C_{js}}{T(T-1)}
 =\frac{S_C}{T-1},
\]
which proves \eqref{eq:H}.
Total variance gives $\E\omega_{NT}^2(Y_{NT})=1-N\VF=(1-N/T)\sC+(1-1/T)\sI$. By \eqref{eq:H} and variance component consistency, $\omega_{NT}^2(Y_{NT})-\E\omega_{NT}^2(Y_{NT})=\smallop(1)$. Consequently, along the chosen subsequence,
\[
 \frac{\omega_{NT}^2(Y_{NT})}{NB\VB}
   =1-\frac{\VF}{\VB}+\smallop(1)
   \ \xrightarrow{\mathrm P}\ \rho:=1-\ell .
\]
If $\rho=0$, the normalized matching error $(\widehat\theta_{NT,B}-\dbar)/\sqrt{\VB}$ vanishes in mean square. If $\rho>0$, \eqref{eq:floor} gives $\Pr(\omega_{NT}^2(Y_{NT})>\rho/2)\to1$.

Consider first $B\to\infty$ with $\rho>0$. Because the $N$ observations in one matching are i.i.d., $\E[(\sqrt N M_{NT,1})^4]\le K/N+3$. As $\dbar=\E[M_{NT,1}\mid Y_{NT}]$, Jensen's and Markov's inequalities yield $\E[N^2D_{NT,1}^4\mid Y_{NT}]=\op(1)$. The conditional Lyapunov ratio is therefore
\[
 \frac{\sum_{b=1}^B \E[D_{NT,b}^4\mid Y_{NT}]}
      {\{\sum_{b=1}^B\Var(D_{NT,b}\mid Y_{NT})\}^2}
 =\frac{\E[N^2D_{NT,1}^4\mid Y_{NT}]}{B(\omega_{NT}^2(Y_{NT}))^2}
 =\smallop(1).
\]
For a fixed number of matchings, apply Theorem~1.1 of \citet{chen2015error} to a single permutation of $C$. With $s^2=N\omega_{NT}^2(Y_{NT})$, its conditional normal approximation satisfies
\[
 \sup_x\big|\Pr(ND_{NT,1}/s\le x\mid Y_{NT})-\Phi(x)\big|
 \le \frac{C_0}{Ts^3}\sum_{i,t\le T}|C_{it}|^3=\smallop(1).
\]
Here $C_0$ is a universal constant. The entries of the first $N$ rows before padding have uniformly bounded third moments. Adding the zero rows leaves the sum of absolute cubes unchanged, and subtracting the column means increases it by at most a constant factor by Jensen's inequality. Hence $T^{-1}\sum |C_{it}|^3=\op(N)$, whereas $s^3/N^{3/2}$ is bounded away from zero in probability. Conditional independence gives the Gaussian approximation for the average of any fixed number of draws.

Every integer sequence $B$ has a further subsequence along which it is constant or diverges. On either subsequence, for every real $u$, the preceding limits give
\[
 \E\!\left[\exp\!\left\{\mathrm{i}u(\widehat\theta_{NT,B}-\dbar)/\sqrt{\VB}\right\}\,\middle|\,Y_{NT}\right]
 \xrightarrow{\mathrm P} \exp(-\rho u^2/2).
\]
The conditional characteristic functions are bounded, so convergence also holds in $L^1$. Multiply by $\exp(\mathrm{i}u\dbar/\sqrt{\VB})$ and take expectations. The limiting characteristic function is $\exp[-(\ell+\rho)u^2/2]=\exp(-u^2/2)$. Equation~\eqref{eq:cons} and Slutsky's theorem give \eqref{eq:clt}, since a continuous limiting distribution gives convergence in Kolmogorov distance. The argument applies to every sequence of laws and thus establishes uniformity.

\paragraph{Uniformity under a marginal fourth-moment bound.}
For $P\in\calQ_{NT,4}(K)$, Jensen's inequality gives $\E \arow{1}^4,\E \bcol{1}^4\le K$. In the variance calculation this gives absolute errors $\op(N^{-1/2})$ for $\hR$ and $\hC$, hence relative contributions $\op(B/\sqrt N)$. For the projection sums the fourth-moment bound becomes
\[
 \frac{1}{\VB^2}\left(\frac{\E \arow{1}^4}{N^3}+\frac{\E \bcol{1}^4}{T^3}\right)
 \le\frac{2K}{N^3\VB^2}\le\frac{2KB^2}{N}.
\]
Both bounds vanish when $B=o(\sqrt N)$. The conditional approximation requires only the marginal fourth moment and consistency of the variance components established above. This proves the claim under the marginal fourth-moment bound. \hfill$\square$

\section{Limiting distributions and a boundary example}\label{app:boundary}

\begin{proof}[Proof of Proposition~\ref{prop:dense}]
Center the observations and set $\sY=1$ by rescaling. Write
$D_{NT,b}=M_{NT,b}-\dbar$ and
$\omega_{NT}^2=N\Var(D_{NT,1}\mid Y_{NT})$ as in \eqref{eq:H}.
Total variance and \eqref{eq:v} give
\[
 \E\omega_{NT}^2=N(\Vone-\VF)
 =NB\VB(1-w_{NT,B}),\qquad NB\VB\ge1 .
\]
The marginal fourth-moment bound implies
$\hC-\sC=o_{\mathrm P}(1)$ and $\hI-\sI=o_{\mathrm P}(1)$ by the
variance-component bounds in Appendix~\ref{app:proof}. These bounds
do not involve $B$. Consequently \eqref{eq:H} gives
$\omega_{NT}^2-\E\omega_{NT}^2=o_{\mathrm P}(1)$ on rectangles as well.

If $w<1$, then $\liminf\E\omega_{NT}^2\ge1-w>0$.
Conditional on the array, the matching errors are i.i.d. For
$B\to\infty$, the bounded conditional fourth moment
$\E[N^2D_{NT,1}^4\mid Y_{NT}]=O_{\mathrm P}(1)$ gives a conditional
Lyapunov ratio of order $O_{\mathrm P}(1/B)$. For fixed $B$, the
single-permutation bound following \eqref{eq:H} gives a conditional
normal approximation to each draw, since $\omega_{NT}^2$ is bounded
away from zero in probability. This argument also covers bounded or
oscillating $B$ by taking subsequences. Since
$\omega_{NT}^2/\E\omega_{NT}^2\to1$ in probability, for every fixed
real $t$ we obtain
\[
 \E\!\left[\exp\!\left\{\mathrm i t
  \frac{\widehat\theta_{NT,B}-\dbar}{\sqrt{\VB}}\right\}
   \,\middle|\,Y_{NT}\right]
 \ \longrightarrow\ \exp\{-t^2(1-w)/2\}
 \quad\text{in }L^1 .
\]
Here convergence in probability implies $L^1$ convergence because the
characteristic functions are bounded. If $w=1$, the same display
holds because the matching error on this scale converges to zero in
mean square. Multiplying the conditional characteristic function by
$\exp\{\mathrm i t(\dbar-\target)/\sqrt{\VB}\}$ and taking expectations
gives
\[
 \E e^{\mathrm i t(\widehat\theta_{NT,B}-\target)/\sqrt{\VB}}
 = e^{-t^2(1-w)/2}\,
   \E e^{\mathrm i t\sqrt{w_{NT,B}}(\dbar-\target)/\sqrt{\VF}}+o(1).
\]
For $w>0$ this gives the stated independent sum. For $w=0$ the
full-array error on the total standard deviation scale vanishes in
mean square, so no full-array limit is required. If the variance
share itself does not converge, the same argument applies along
each subsequence on which it does.

To establish studentization when $w=0$, let
$\widehat V_{NT,\mathrm F}=\hR/N+\hC/T+\hI/(NT)$.
Centering the row means gives
$\E\hR=(1-1/N)(\sR+\sI/T)$; centering the column means gives
$\E\hC=(1-1/T)(\sC+\sI/N)$. Double demeaning the residuals gives
$\E\hI\le\sI$. Adding the three contributions gives
$\E\widehat V_{NT,\mathrm F}\le \sR/N+\sC/T+3\sI/(NT)\le3\VF$.
Consequently, for every $\epsilon>0$,
\[
 \Pr\!\left(\widehat V_{NT,\mathrm F}/\VB>\epsilon\right)
 \le \frac{3\VF}{\epsilon\VB}
 = \frac{3w_{NT,B}}{\epsilon}\longrightarrow0 .
\]
By Appendix~\ref{app:proof}, $\hY/\sY\to1$ in probability. Since
\[
 \frac{\sY}{NB\VB}=1-(1-1/B)w_{NT,B}\to1,
\]
\eqref{eq:se} gives $\widehat V_{NT,B}/\VB\to1$ in probability.
\end{proof}

\paragraph{Scales in Table~\ref{tab:regimes}.}
If $\nuNT\to c<1$, Section~\ref{sec:Blimits} gives $\Vone/(T\VF)\to1-c$.
Using $\Vone=\sY/N$ in \eqref{eq:v}, we obtain
\[
 \frac{NT\VF}{\sY}\longrightarrow\frac{1}{1-c},
 \qquad
 \frac{NT(\VB-\VF)}{\sY}
 =\frac{T}{B}\left(1-\frac{N\VF}{\sY}\right).
\]
Thus for $B/T\to\lambda\in(0,\infty)$ the two standard deviations on the $\sqrt{NT}/\sigma_{NT}$ scale tend to $1/\sqrt{1-c}$ and $1/\sqrt{\lambda}$, respectively. For $B/T\to\infty$ the second tends to zero. For $B=o(T)$ use the $\sqrt{NB}/\sigma_{NT}$ scale instead: $NB\VF/\sY\to0$ and $NB\VB/\sY=1+(B-1)N\VF/\sY\to1$.

\paragraph{When the projections dominate.}
Let $V_{NT,\mathrm P}=\sR/N+\sC/T=\nuNT\VF$ and suppose $\nuNT\to1$.
The intersection average divided by $\sqrt{\VF}$ has variance
$1-\nuNT\to0$. The independent projection summands enter the mean with weights $1/N$ and $1/T$. Their Lindeberg condition requires, for every $\epsilon>0$,
\[
 \frac{\E[\arow{1}^2\mathbf 1\{|\arow{1}|>\epsilon N\sqrt{V_{NT,\mathrm P}}\}]}{N V_{NT,\mathrm P}}
 +\frac{\E[\bcol{1}^2\mathbf 1\{|\bcol{1}|>\epsilon T\sqrt{V_{NT,\mathrm P}}\}]}{T V_{NT,\mathrm P}}
 \longrightarrow 0 .
\]
Under this condition the normalized projection sum is asymptotically standard normal. Since $V_{NT,\mathrm P}/\VF\to1$ and the residual average is negligible on this scale, the same limit holds for the standardized full-array mean. In particular, for laws in
$\calP_{NT}(K)$, the Lyapunov ratio satisfies
\[
 \frac{\E[\arow{1}^{4}]/N^3+\E[\bcol{1}^{4}]/T^3}{V_{NT,\mathrm P}^2}
 \le \frac{K\sY}{N^2 V_{NT,\mathrm P}}
 \le K\frac{T}{N}\left(\frac1N+\frac{1-\nuNT}{\nuNT}\right).
\]
This tends to zero if $T/N$ is bounded. Without such a restriction
the conclusion need not hold. For example, take $T=N^2$ and
$Y_{NT,it}=I_{NT,i}-1/N+\zeta_{it}$, with independent
Bernoulli$(1/N)$ row shocks and independent standard normal
intersection shocks, independent of the row shocks. Here
$\sR=(1-1/N)/N$, $\sC=0$, and $\sI=1$. The moment bounds defining
$\calP_{NT}(K)$ hold for a common $K$, since the row projection is
bounded by one and $\sY\ge1$. Yet $N^2\VF=1$ and
$\nuNT=1-1/N\to1$, while $N\dbar\Rightarrow\mathrm{Poisson}(1)-1$:
the row count converges to a Poisson law and the intersection
average has variance $1/N$ on this scale.

\paragraph{Sharpness under a marginal fourth-moment bound.}
Suppose $B=o(N)$ and $B/\sqrt N$ does not tend to zero. Work along a subsequence on which $B\ge c\sqrt N$ for some $c>0$. For these array dimensions $N\le T$, let
\[
 Y_{NT,it}=\arow{i}+e_{it},\qquad
 \arow{i}=\sqrt{N/B}\,(I_{NT,i}-1/N),
 \qquad I_{NT,i}\overset{\mathrm{i.i.d.}}{\sim}\operatorname{Bernoulli}(1/N),
\]
where the $e_{it}$ are independent standard normals independent of the $I_{NT,i}$. Here $\target=0$, $\sR=(1-1/N)/B$, $\sC=0$, $\sI=1$, and $\E[\arow{i}^4]\sim N/B^2=O(1)$. Hence the standardized marginal fourth moment is bounded, while $\E[\arow{i}^4]/(\sY\sR)\sim N/B\to\infty$. Each matching uses every row once, so the row contribution is
\[
 \frac{\sqrt{NB}}{N}\sum_i\arow{i}=\sum_{i=1}^N I_{NT,i}-1
       \ \Rightarrow\ Q-1,\qquad Q\sim\operatorname{Poisson}(1).
\]
Let $K_{NT,it}$ count the number of times intersection $(i,t)$ is
selected. For any row $i$ and distinct matchings $b,b'$, the assigned
columns coincide with probability $1/T$. Summing over $N$ rows and
$B(B-1)$ ordered pairs of matchings gives
\[
 \E\sum_{i,t}K_{NT,it}(K_{NT,it}-1)=\frac{NB(B-1)}{T}.
\]
Since $\sum_{i,t}K_{NT,it}=NB$ and $B=o(N)$ with $N\le T$,
\[
 \E\left[\frac{\sum_{i,t}K_{NT,it}^2}{NB}-1\right]
 =\frac{B-1}{T}\to0.
\]
The quantity in brackets is nonnegative, so
$\sum_{i,t}K_{NT,it}^2/(NB)\to1$ in probability by Markov's inequality.
Conditional on the matchings, the intersection error in $\sqrt{NB}\,\widehat\theta_{NT,B}$ is Gaussian with this variance and is independent of the row projections. Thus $\sqrt{NB}\,\widehat\theta_{NT,B}\Rightarrow Q-1+Z$, where $Z$ is standard normal and independent of $Q$.

The variance estimate $\widehat V_{NT,B}$ is positive almost surely, so the studentized statistic has the same sign as $\widehat\theta_{NT,B}$. Its distribution function at zero converges to
\[
 \E\Phi(1-Q)
 \ \ge\ e^{-1}\{\Phi(1)+\tfrac12+\tfrac12\Phi(-1)\}
 \ >\ 0.52 .
\]
Since $\Phi(0)=1/2$, a marginal fourth-moment bound alone cannot guarantee uniform Gaussian approximation for these choices of $B$. This includes the boundary choice $B=\lfloor\sqrt N\rfloor$.

\section{Proof of Theorem~\ref{th:gmm}}\label{app:gmm}
Work along any admissible sequence of laws. Applying \eqref{eq:v} to a scalar function $x$ of an intersection observation bounds the variance of its matching average by $\Var(x_{NT,11})/N$. At $\btrue$, this bound applied componentwise gives
\[
 \widehat{\bm m}_{NT,B}(\btrue)=O_{\mathrm P}(N^{-1/2}),\qquad
 \partial\widehat{\bm m}_{NT,B}(\btrue)/\partial\bm\beta^\top
                  -\bm D_{NT}=O_{\mathrm P}(N^{-1/2}).
\]
Applied on a finite net of $\Theta$, the same bound gives convergence to the population moments on that net. The sampled and population averages of the derivative envelope are $O_{\mathrm P}(1)$ and $O(1)$, respectively. Taking finer nets therefore gives uniform convergence over $\Theta$. The weight bounds and uniform identification imply $\widehat{\bm\beta}_{NT,B}-\btrue=o_{\mathrm P}(1)$, so the minimizer is interior with probability tending to one.

Write $\bm\Delta_{NT,B}=\widehat{\bm\beta}_{NT,B}-\btrue$, and let $\overline{\bm D}_{NT,B}$ be the average sampled derivative along the segment joining these parameter values. The second-derivative bound and consistency give $\widehat{\bm D}_{NT,B}-\bm D_{NT}=o_{\mathrm P}(1)$ and $\overline{\bm D}_{NT,B}-\bm D_{NT}=o_{\mathrm P}(1)$. With probability tending to one, the solution is interior and the derivative product below is invertible. Substitution in the first-order condition gives
\begin{align*}
 \bm 0&=\widehat{\bm D}_{NT,B}^{\top}\widehat{\bm W}_{NT}
 \bigl\{\widehat{\bm m}_{NT,B}(\btrue)+\overline{\bm D}_{NT,B}\bm\Delta_{NT,B}\bigr\},\\
 \bm\Delta_{NT,B}
 &=-\widetilde{\bm A}_{NT,B}\,\widehat{\bm m}_{NT,B}(\btrue)
 =O_{\mathrm P}(N^{-1/2}),
\end{align*}
where $\widetilde{\bm A}_{NT,B}=(\widehat{\bm D}_{NT,B}^{\top}\widehat{\bm W}_{NT}\overline{\bm D}_{NT,B})^{-1}
\widehat{\bm D}_{NT,B}^{\top}\widehat{\bm W}_{NT}$. The derivative errors are now $O_{\mathrm P}(N^{-1/2})$, so
$\widetilde{\bm A}_{NT,B}-\bm A_{NT}=O_{\mathrm P}(N^{-1/2})$. A Taylor expansion of $\ell$ and the definition of $u_{NT,it}$ in \eqref{eq:gmm-score} give
\begin{align*}
 \ell(\widehat{\bm\beta}_{NT,B})-\ell(\btrue)
 &=\frac{1}{NB}\sum_{b,i}u_{NT,i\,\pi_b(i)}+R_{NT,B},\\
 |R_{NT,B}|
 &\le C\left(\|\bm\Delta_{NT,B}\|^2
  +\|\widetilde{\bm A}_{NT,B}-\bm A_{NT}\|\,
    \|\widehat{\bm m}_{NT,B}(\btrue)\|\right)
 =O_{\mathrm P}(N^{-1}).
\end{align*}
By \eqref{eq:v}, $\VB(u_{NT})\ge\Var(u_{NT,11})/(NB)\ge\underline v/(NB)$.
Thus $R_{NT,B}/\sqrt{\VB(u_{NT})}=O_{\mathrm P}(\sqrt{B/N})=o_{\mathrm P}(1)$.

For feasible studentization, the full-array average of $(\widehat u_{NT,it}-u_{NT,it})^2$ is $O_{\mathrm P}(N^{-1})$. Indeed, the coefficient vector in \eqref{eq:gmm-score} has estimation error $O_{\mathrm P}(N^{-1/2})$, and the full-array average of the squared envelope is $O_{\mathrm P}(1)$. Formula~\eqref{eq:se} is a sum of squared linear functions with nonnegative coefficients, so its square root is a seminorm. The row, column, and residual sums of squares add to the centered marginal sum of squares. With $N\le T$, this gives $\widehat V_{NT,B}(x)\le N^{-1}(NT)^{-1}\sum_{i,t}x_{NT,it}^2$. Hence
\begin{align*}
 \left|\sqrt{\widehat V_{NT,B}(\widehat u_{NT})}
       -\sqrt{\widehat V_{NT,B}(u_{NT})}\right|
 &\le \sqrt{\widehat V_{NT,B}(\widehat u_{NT}-u_{NT})}\\
 &\le \left(\frac{1}{N^2T}\sum_{i,t}(\widehat u_{NT,it}-u_{NT,it})^2\right)^{1/2}
 =O_{\mathrm P}(N^{-1}).
\end{align*}
Applying Theorem~\ref{th:main} to $u_{NT}$ proves \eqref{eq:gmm-clt}, including the assertion under $B=o(\sqrt N)$.

\paragraph{Joint inference.}
For a unit vector $\bm c$, take $\ell(\bm\beta)=\bm c^\top\bm\beta$ and denote its influence contribution by $u_{NT}(\bm c)$. Suppose the variance and moment bounds of Theorem~\ref{th:gmm} hold with common constants for all such $\bm c$. Define covariance matrices $\bm S_{NT,B}$ and $\widehat{\bm S}_{NT,B}$ by the quadratic forms $\bm c^\top\bm S_{NT,B}\bm c=\VB(u_{NT}(\bm c))$ and $\bm c^\top\widehat{\bm S}_{NT,B}\bm c=\widehat V_{NT,B}(\widehat u_{NT}(\bm c))$. Since the preceding argument applies to every deterministic sequence of unit directions, Cram\'er-Wold applied in directions proportional to $\bm S_{NT,B}^{-1/2}\bm c$ gives $\bm S_{NT,B}^{-1/2}(\widehat{\bm\beta}_{NT,B}-\btrue)\Rightarrow \mathcal N(\bm 0,\bm I_d)$. Directional variance consistency and polarization on the same scale give $\bm S_{NT,B}^{-1/2}\widehat{\bm S}_{NT,B}\bm S_{NT,B}^{-1/2}\to\bm I_d$ in probability. Thus the inverse exists with probability tending to one and $(\widehat{\bm\beta}_{NT,B}-\btrue)^\top\widehat{\bm S}_{NT,B}^{-1}(\widehat{\bm\beta}_{NT,B}-\btrue)\Rightarrow\chi^2_d$. The arbitrary-sequence argument establishes uniformity. \hfill$\square$

\clearpage
\bibliographystyle{chicago}
{\small\setstretch{1.0}\bibliography{references}}
\end{document}